\PassOptionsToPackage{disable}{lineno}
\documentclass[9pt,twocolumn,twoside]{pnas-new}
\templatetype{pnasresearcharticle} % Choose template
\begin{document}

\title{A minimal wake-vortex model explains formation flight of flapping birds}

% Use letters for affiliations, numbers to show equal authorship (if applicable) and to indicate the corresponding author
\author[a,1]{Olivia Pomerenk}
\author[a]{Kenneth S. Breuer}

\affil[a]{Center for Fluid Mechanics, School of Engineering, Brown University, Providence, RI 02912, USA}

% Please give the surname of the lead author for the running footer
\leadauthor{Pomerenk}

% Please add a significance statement to explain the relevance of your work
\significancestatement{The striking formation patterns of flocking birds have long been hypothesized to yield energetic benefits, which arise from unsteady aerodynamic interactions between birds' wings and their neighbors' vortical wakes. Yet despite decades of research, the mechanisms underlying energy savings in formation flight remain poorly understood. Past models either neglect the inherent unsteadiness of flapping flight, oversimplifying the essential physics, or they directly simulate the full flow field, obscuring minimal mechanisms. In this work, we propose a tractable middle-ground analytical model that captures unsteady flight dynamics without direct numerical simulation. This generalizable framework clarifies the specific mechanisms by which flapping birds exploit their leaders' vortical wakes, and it provides a foundation for future studies of collective flight.}

% Please include corresponding author, author contribution and author declaration information
\authorcontributions{O.P.: conceptualization, methodology, analysis, validation, writing original draft, funding acquisition. K.B.: conceptualization, supervision, review and editing, funding acquisition. }
\authordeclaration{The authors have no competing interests to declare.}
% \equalauthors{\textsuperscript{1}A.O.(Author One) contributed equally to this work with A.T. (Author Two) (remove if not applicable).}
\correspondingauthor{\textsuperscript{1}To whom correspondence should be addressed. E-mail: olivia\_pomerenk@brown.edu}

% At least three keywords are required at submission. Please provide three to five keywords, separated by the pipe symbol.

\begin{abstract} % 250 words

Collective patterns of motion emerge across biological taxa: insects swarm, fish school, and birds flock. In particular, many large migratory bird species form distinctly ordered V-shaped formations, which experiments and direct numerical simulations have demonstrated provide substantial energetic benefits during long-distance flight. However, the precise aerodynamic and morphological mechanisms which underlie these benefits remain unclear. In this work, we develop a reduced-order model of the wake-vortex interactions between two flapping birds flying in tandem. The model retains essential unsteady flapping dynamics while remaining computationally tractable. By optimizing over a six-dimensional state space, which comprises the follower's three-dimensional relative position as well as three independent flapping parameters, we identify the energetically optimal leader-follower configuration of northern bald ibises (\textit{Geronticus eremita}). The predicted optimum agrees quantitatively with live-bird measurements. Because of its simplicity, the model allows for direct interrogation of the physical mechanisms responsible for this optimum. In particular, it isolates precisely how the follower's wing kinematics interact with the leader's wake to enhance aerodynamic efficiency. The model predicts an 11\% reduction in total mechanical power for a follower in formation flight -- consistent with experimental estimates -- and shows that this saving arises from reductions in both induced and profile power, dominated by decreased profile power enabled primarily through reduced flapping amplitude and, secondarily, reduced upstroke flexion. These results provide a mechanistic explanation for the structure of V-formations and offer new insight into the aerodynamic principles governing collective flight.

\end{abstract}

\dates{This manuscript was compiled on \today}
\doi{\url{www.pnas.org/cgi/doi/10.1073/pnas.XXXXXXXXXX}}

\maketitle
\thispagestyle{firststyle}
\ifthenelse{\boolean{shortarticle}}{\ifthenelse{\boolean{singlecolumn}}{\abscontentformatted}{\abscontent}}{}

\firstpage[6]{2}

\dropcap{M}any species of large migratory birds, such as ibises or geese, have long been observed to fly in distinctive V-shaped configurations \cite{lissaman1970formation,hummel1983aerodynamic,portugal2014upwash,friman2024pays,beaumont2024aerodynamic,hainsworth1988induced, corcoran2019compound}. While real avian flocks -- in which individuals continually adjust position, exchange roles, and exit and re-enter the group formation -- are highly dynamic \cite{voelkl2015matching}, analysis of the idealized static V-formation serves as a baseline with which to parse the more complex dynamics of group flight \cite{lissaman1970formation, friman2024pays}. Vortical wake interactions among such flapping birds are broadly hypothesized to be the primary mediator of these flocking patterns \cite{newbolt2024flow}, and this hypothesis is strongly supported by experimental live-bird studies \cite{portugal2014upwash,friman2024pays,hedenstrom2006vortex} and computational fluid dynamic (CFD) simulations \cite{beaumont2022modeling,beaumont2024aerodynamic}. It is generally understood that organized formations offer significant aerodynamic advantages to avian collections, the most significant of which involves a reduction in members' energy expenditure during flight \cite{weimerskirch2001energy, hainsworth1988induced,cutts1994energy,voelkl2015matching, hummel1983aerodynamic}. The mechanism for these energetic savings is broadly understood to work as follows \cite{lissaman1970formation,hummel1983aerodynamic}: a flapping ``leader'' bird sheds a pair of counter-rotating vortices from its wingtips to its wake, and these tip vortices produce regions of so-called upwash (upward flow) in the exterior spanwise region of the leader and regions of downwash (downward flow) in the interior spanwise region. A downstream ``follower'' bird flies in this aerodynamically beneficial region of upwash, which reduces the follower's energetic requirements for flight. Followers therefore tend to select their average position such that they fly near their immediate leader's upwash region -- that is, behind and to the side of their leader. Although in real flight, birds dynamically exchange their positioning within the flock according to myriad physical and social cues -- which is itself an active area of research \cite{voelkl2015matching, fernandez2004visual, salahshour2025allocentric} -- it is generally held that this upwash/downwash effect broadly gives rise to the characteristic V-formation associated with myriad bird species.

However, this prevailing upwash/downwash argument is highly simplified to the point of being incomplete, even when applied to the reduced-order static configuration setting. Real flapping birds generate unsteady vortical wake patterns with extremely complex geometries, and both the forms of these vortex structures themselves, as well as their interactions with a dynamically flapping follower bird, remain active areas of research \cite{hedenstrom2006vortex, spedding2003family, beaumont2024aerodynamic}. Reports disagree on whether the aerodynamic benefit of upwash involves a reduction of induced drag, an increase in lift, or some combination of the two \cite{voelkl2015matching,hummel1983aerodynamic,hainsworth1988induced,portugal2014upwash}. Moreover, existing studies fail to encapsulate key dynamical aspects of fixed-formation flight: for example, there does not exist any model which successfully accounts for the well-documented temporal flapping phase offsets of birds flying in V-formations \cite{portugal2014upwash}. Thus, while existing experiments \cite{portugal2014upwash,friman2024pays,hedenstrom2006vortex, weimerskirch2001energy,voelkl2015matching} and computational simulations \cite{beaumont2022modeling,beaumont2024aerodynamic, willis2007computational} have reported the salient emergent behaviors of flocking birds to high fidelity, such studies are limited to being primarily observational rather than explanatory -- and they are unable to determine or concisely express the precise underlying mechanisms which govern coherent flock structures. The aerodynamic source of the benefit (i.e., upwash) is understood reasonably well, but the morphological and kinematic realization at the level of the bird itself remains incompletely characterized.

Highly simplified models which reduce dynamically flapping birds to static, fixed-wing structures have also been developed to augment these experimental and numerical studies \cite{hummel1983aerodynamic,lissaman1970formation}. While these models have been historically useful in characterizing the simple upwash/downwash argument described above, they lie at the other end of the proverbial spectrum of complexity as their experimental and computational counterparts: fixed-wing models leave entirely unaddressed the unsteadiness and three-dimensionality inherent in the problem of flocking birds, and are thus too simple to describe the salient physics. As noted by a recent review \citep{beaumont2024aerodynamic}, existing models are either too static (fixed wing, ignore flapping) or too detailed (full numerical simulation). Neither class of model is equipped to give a compact description of the minimal rules governing flapping formation flight.

There does not yet exist a theoretical model which combines just the essential degrees of freedom -- e.g., wingbeat phase, lateral and longitudinal offset, vertical spacing, flapping amplitude, etc. -- without resorting to explicitly resolving the three-dimensional fluid flows, and which reproduces the salient phenomena observed during flocking flight. Such a model would account for the unsteady flapping aspects of the problem while still remaining fundamentally simple, and would thus reveal the essential aerodynamic mechanisms which underlie energy savings for flapping birds in collective formations. Here we pursue the development of such a model.

\section*{Construction of time-averaged force model}

We construct a minimal, theoretical model of the time-averaged forces incident on a pair of flapping birds throughout a stroke cycle. The scope of the model is restricted to considering a fixed-position formation and thus neglects several aspects of real avian flocking, in which birds continually re-organize and exchange their positioning within a flock. Throughout this work, we neglect all aspects of birds' anatomy other than their wings. Birds are assumed to fly at constant speed in an arbitrary three-dimensional spatial configuration relative to each other. To counter gravity and drag (both assumed constant over time), each bird generates lift and thrust by flapping its wings. This flapping imparts an unsteady trailing vortical wake downstream of each bird. Assuming that the birds do not fly precisely side-by-side, the upstream bird (hereafter the ``leader'') is unaffected by the vortical wake of the downstream bird (the ``follower''), whereas the follower may interact with the vortical wake of the leader and thereby experience an external force in addition to its own flapping-supplied force. Thus, the interaction is non-reciprocal insofar as the leader affects the follower, but not vice versa. 

In modeling the aerodynamic interaction of a single pair of birds during a single stroke cycle for a fixed relative position and flight speed, the model presented here may be viewed as a ``kernel'' of a fully-fledged group flight model: it provides an instantaneous snapshot of a pair of birds flying together, rather than describing the entire dynamics over the course of a flight. Future work can reconstruct a more realistic flight history by allowing these, and other, parameters to vary over time, and repeatedly applying the model presented here.

\subsection*{Net force generated by a lone flapping bird}

We first develop a simple model of the time-averaged forces generated by a lone bird flying through quiescent fluid with uniform horizontal speed $U$ and fluid density $\rho$. We set our frame of reference fixed with this bird so that the free-stream speed is $U$. This construction is adapted directly from \cite{spedding2003family}. 

Flapping birds' wakes comprise time-dependent vortical structures which are shed downstream from the wings and whose geometry is generally complex and highly three-dimensional \cite{spedding2003family,rayner1979vortex,parslew2013theoretical,beaumont2022modeling,beaumont2024aerodynamic}. For a large bird traveling at high speed, the wake roughly resembles two long vortex tubes which oscillate in the vertical direction. As the bird sweeps its wings down during the first phase of its wingbeat, the resultant wake structure follows the wings and is angled downward in the vertical direction. Throughout the downstroke, the birds' wings are fully extended in the spanwise direction, and the corresponding tip vortices are maximally separated. During the upstroke, by contrast, the wake is angled upward in the vertical direction, and the tip vortices move closer together as the bird's wings contract to effectively reduce their span. At the juncture of these upstroke- and downstroke-generated wake segments, starting and stopping vortical structures form in the spanwise direction. As a result, the overall wake resembles a series of closed or semi-closed vortex rings.

\begin{figure}%[tbhp]
\centering
\includegraphics[width=\linewidth]{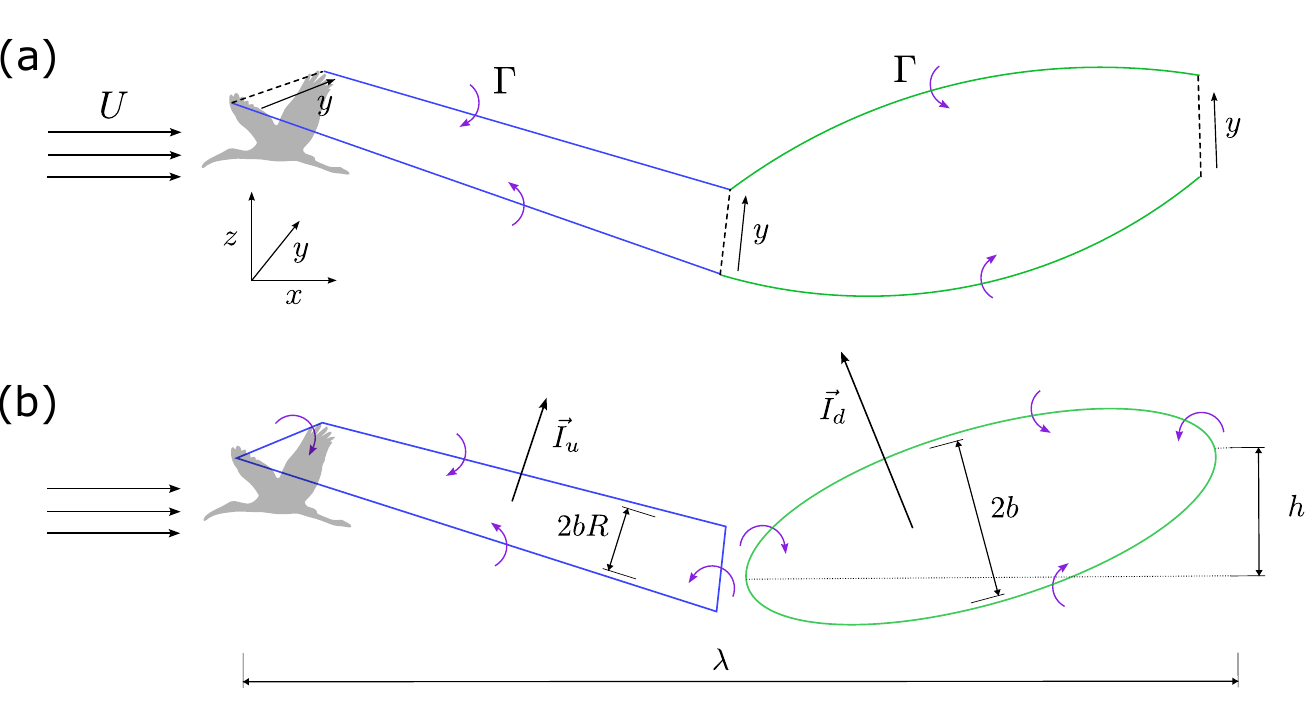}
\caption{Constant-circulation vortical wake of a flapping bird at the end of its upstroke, adapted directly from Spedding et al. \cite{spedding2003family}. (a) Approximation of true wake structure: wings sweep out on the downstroke, forming curved filaments (green), and contract during the upstroke, forming straight filaments (blue). This contraction is modulated by the upstroke flexion ratio $R\in(0,1)$. Cross-stream vortices in the spanwise direction $y$, denoted by dashed lines, have zero circulation. (b) A simplified version of (a), again adapted directly from \cite{spedding2003family}, constructed to succinctly express upstroke and downstroke impulses mathematically. The wake is modeled as a series of vortex filaments, which form closed rectangles and ellipses. The resulting impulse $\vec{I}$ always points upward in the $z$ direction, but due to wingspan contraction on the upstroke, the projected area of the elliptical downstroke region is larger, which generates a net propulsive impulse in the $-x$ (upstream) direction.}
\label{fig:leader_impulse}
\end{figure}

The wake of a lone flapping bird may thus be approximated by a pair of undulating trailing tip vortices of equal and opposite circulation $\pm\Gamma$ (Fig. \ref{fig:leader_impulse}(a)). In real avian flight, the upstroke- and downstroke-generated wake circulations may differ, with $\Gamma_{\rm up} \leq \Gamma_{\rm down}$. As flight speed $U$ increases, $\Gamma_{\rm up}\to\Gamma_{\rm down}$ \cite{spedding2003family}. In this work, which presents results for a fast-flying avian species (\textit{Geronticus eremita}), we approximate $\Gamma_{\rm up} = \Gamma_{\rm down} = \Gamma$ due to a lack of empirical data of $\Gamma_{\rm up}$. However, the formulation allows for arbitrary $\Gamma_{\rm up}$ and $\Gamma_{\rm down}$, and so this assumption can be easily modified.

The time-averaged force generated by such a bird over a single flapping period may be expressed by considering the net impulse associated with a simplified elliptical-rectangular vortex wake geometry \cite{spedding2003family}, as shown schematically in Fig. \ref{fig:leader_impulse}(b). The bird's average vertical force over one wake period may be written in terms of its total vertical impulse $I_z$ and flapping period $T$:
\begin{equation}\label{leader_vert_force}
    F_z = \frac{I_z}{T} \quad\text{where} \quad I_z = I_{z,d} + I_{z,u}.
\end{equation}
Here, the total impulse has been decomposed into its components generated by the downstroke ($d$) and upstroke ($u$). These components are respectively given by
\begin{equation}\label{leader_vert_imp}
    I_{z,d} = \frac{\rho\pi b\lambda\Gamma}{4} \quad\text{and}\quad 
    I_{z,u} = \rho b R \lambda\Gamma.
\end{equation}
In the above, $\rho$ is the density of the fluid; $b$ is the bird semi-span; $\lambda=UT$ is the wavelength in the horizontal ($x$) direction of a full wake (combined downstroke and upstroke); $\Gamma$ is the magnitude of the constant circulation associated with wake vortex filaments; and $R\in (0,1)$ is the projected relative span ratio of the bird's wing during its upstroke. These different forms of $I_{z,d}$ and $I_{z,u}$ are due to the elliptical vs. rectangular projected span areas in the $(x,y)$ plane. Here we have implicitly assumed that the proportion of the total wingbeat cycle spent in the downstroke is equal to that of the upstroke, which aligns with observations and modeling approaches taken in past works \cite{tobalske1996flight,portugal2014upwash}. This construction is summarized by Fig. \ref{fig:leader_impulse}(b).

Similar accounting is done for the force in the horizontal direction, which is given by
\begin{equation}
    F_x = \frac{I_x}{T}\quad\text{where}\quad I_x = I_{x,d} - I_{x,u},
\end{equation}
with
\begin{equation}\label{leader_hor_force}
    I_{x,d} = \frac{\rho\pi b h\Gamma}{2}\quad\text{and}\quad
    I_{x,u} = 2\rho b R h\Gamma.
\end{equation}
Here, $h$ is the tip-to-tip flapping amplitude of the bird.

\subsection*{Net forces on a pair of birds flying in tandem}

With the above formulation of the forces generated by a single flapping bird, we may now consider a leader-follower pair of birds. The quantities $\rho$, $b$, $\lambda$, $T$, and $\Gamma$ are assumed to be constant for both birds -- reflecting similar bird geometry, weight and flight speed -- while the tip-to-tip flapping amplitude $h$ and upstroke flexion ratio $R$ may be unequal, $h_L\neq h_F$ and $R_L\neq R_F$ \citep{spedding2003family,portugal2014upwash,tobalske1996flight}.

The leader bird flies as though alone: it does not interact with the trailing wake structure of the follower. The leader's total lift and thrust are therefore given directly by Eqs. (\ref{leader_vert_force} - \ref{leader_hor_force}) with $h=h_L$ and $R=R_L$.

To model the time-averaged forces for a follower bird, however, we must consider both the force that the follower itself generates via flapping as well as the fluid dynamical force induced by the unsteady trailing wake vortices of the leader with which the follower interacts. We write the total vertical and horizontal force on the follower as
\begin{equation}
    F_z^F = \frac{I_z^F}{T} + F_z^I\quad\text{and}\quad  
    F_x^F = \frac{I_x^F}{T} + F_x^I. \label{total_follower_force}
\end{equation}
Here, the terms $I_{z,x}^F/T$ correspond to the follower's flapping-generated force (Eqs. (\ref{leader_vert_force} - \ref{leader_hor_force}), with $h=h_F$ and $R=R_F$), while the terms $F_{z,x}^I$ correspond to the additional force generated by interaction with the leader's vortical wake. We now define forms for these wake-induced forces $F_z^I$ and $F_x^I$.

A central complication of this problem involves the fundamentally unsteady nature of both the leader's wake and the follower's own flapping. To deal with this unsteadiness in a minimal fashion, we introduce a major simplifying assumption: at any time $t\in[0,T]$ during a flapping cycle, we assume that both the leader and follower are each in precisely one of two possible static configurations. These are ``up,'' i.e., at the end of the upstroke, or ``down,'' i.e., at the end of the downstroke. Thus, there are four leader-follower possible configurations: up-up, up-down, down-up, and down-down. This nontrivial assumption is made to maintain the model's simplicity and tractability.

\begin{figure}[t!]
\centering
\includegraphics[width=\linewidth]{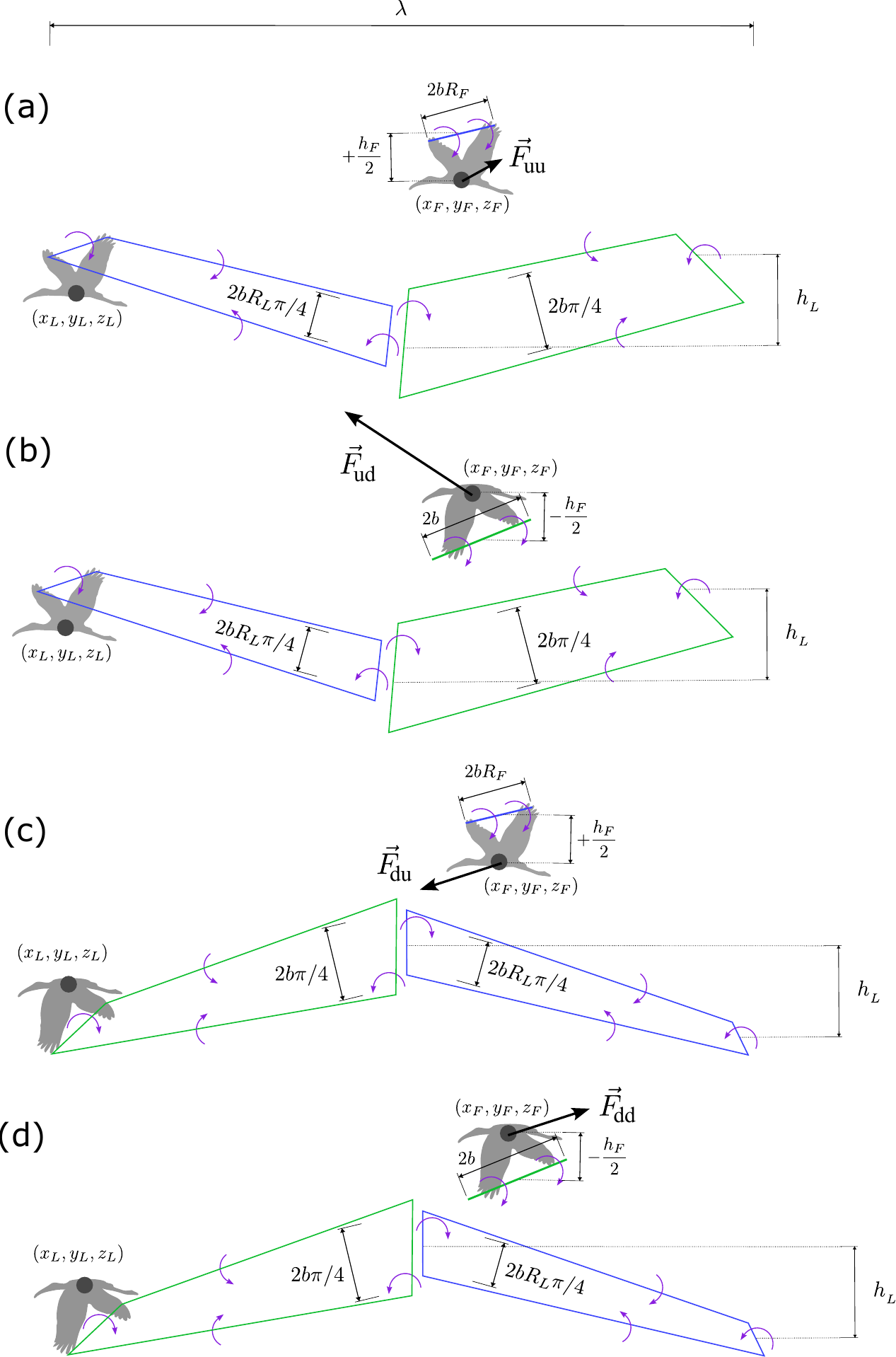}
\caption{Available leader-follower configurations: up-up (a), up-down (b), down-up (c), and down-down (d). Rankine vortex filaments, which comprise an approximation of the leader's unsteady vortical wake, are indicated by straight green (downstroke) and blue (upstroke) lines. The leader's wake is a simplified version of that in Fig. \ref{fig:leader_impulse}(b): here, the elliptical downstroke wake portion is reduced to a rectangle, akin to the upstroke portion, for simplicity. Both the upstroke and downstroke rectangle widths are modified by a wingtip vortex roll-up factor of $\pi/4$ \cite{hummel1983aerodynamic, hainsworth1988induced}. Vortices are endowed with signed constant circulation $\pm\Gamma$ (purple). The follower is modeled as a lifting line directed along the spanwise $y$-axis, with symmetric circulation distribution such the circulation at each endpoint is equal to $\Gamma$. The net wake-induced force on the follower ($\vec{F}_{\rm{uu}}$, $\vec{F}_{\rm{ud}}$, and so on) is marked by a black arrow in each panel (arrow not to scale). The follower ``flaps'' by heaving up and down in the $z$ direction with amplitude $h_F/2$.}
\label{fig:wake_structure}
\end{figure}

The leader's up/down configuration is realized by its vortical wake structure, which emanates from the leader's fixed reference coordinates $(x_L,y_L,z_L)$. We simplify the elliptical-rectangular wake model used in the earlier impulse-based force formulation by reducing the elliptical downstroke portion of the wake to a rectangle of width $2b$. Figs. \ref{fig:wake_structure}(a,b) depict the leader-up wake structure, while Figs. \ref{fig:wake_structure}(c,d) depict the leader-down wake structure. Over one flapping cycle, the leader's wake alternates between these two discrete, fully developed states. The upstroke configuration (Figs. \ref{fig:wake_structure}(a,b)) persists from $t=0$ to $t=T/2$, and the downstroke configuration (Figs. \ref{fig:wake_structure}(c,d)) persists from $t=T/2$ to $t=T$. To construct a given wake configuration, we introduce eight finite-length vortex filaments which together form two rectangles: one associated with the leader's upstroke wake, and the other with the leader's downstroke wake. Each filament is endowed with signed circulation $\pm\Gamma$. Right-hand-side (RHS) filaments and starting vortices are associated with positive circulation, while left-hand-side (LHS) filaments and stopping vortices are associated with negative circulation. To account for viscosity and avoid discontinuities, these vortex filaments are endowed with Rankine cores, i.e., cores with solid-body fluid rotation \cite{Young2003VortexCore}.

Meanwhile, the follower acts independently of the leader, and may itself execute either an upstroke (Figs. \ref{fig:wake_structure}(a,c)) or a downstroke (Figs. \ref{fig:wake_structure}(b,d)). The follower is modeled as a straight lifting-line vortex filament directed along the spanwise $y$-axis. The follower's downstroke is characterized by the entire lifting line being displaced in the negative vertical direction by $h_F/2$, and the upstroke with displacement upward by $h_F/2$. Thus, the follower's wing discretely heaves up and down, and its $z$ coordinate is $z_F+h_F/2$ or $z_F-h_F/2$ during its ``up'' and ``down'' configurations, respectively. This discrete ``flapping'' of the follower is illustrated in Fig. \ref{fig:wake_structure}.

Each of these 4 leader-follower configurations is associated with a wake-induced force on the follower: $\vec{F}_{\rm{uu}}$ corresponds to the induced force during the up-up configuration, and so on. To compute each force, we invoke a form of the Biot-Savart law for finite-length Rankine vortex filaments \cite{bertin1998aerodynamics} to calculate the flow velocity induced at each endpoint of the follower wing. We then use a vector form of the Kutta-Joukowski theorem \cite{cummings2015applied,bai2014generalized} to calculate the resultant magnitude and direction of the induced force contribution from each finite-length Rankine vortex filament. Details of this procedure are provided as Materials and Methods.

To compute the total time-averaged wake-induced force on the follower, the four forces associated with the available leader-follower configurations ($\vec{F}_{\rm{uu}}$, $\vec{F}_{\rm{ud}}$, $\vec{F}_{\rm{du}}$, and $\vec{F}_{\rm{dd}}$) are each assigned a weight $\beta$. This weighting is determined by the temporal phase offset of the follower's flapping relative to its leader.

As the follower is not restricted to flapping in temporal phase with the leader, it may displace the beginning of its upstroke to occur at any time $t=\kappa_F T$ where $\kappa_F\in[0,1)$. Thus, $\kappa$ is the proportion of a flap cycle of the leader at which the follower initiates a flap. Note that the leader is necessarily associated with $\kappa_L=0$. It can be shown that this yields temporal weights for the down-down and up-up configurations:
\begin{equation}\label{eqn:beta_uniform}
    \beta_{\rm{dd}} = \beta_{\rm{uu}} = \biggr\rvert \kappa_F-\frac{1}{2}\biggr\rvert,
\end{equation}
and for the down-up and up-down configurations:
\begin{equation}\label{eqn:beta_mixed}
    \beta_{\rm{du}} = \beta_{\rm{ud}} = \frac{1}{2}-\biggr\rvert \kappa_F-\frac{1}{2}\biggr\rvert.
\end{equation}
Here, $\beta_{\rm{dd}}$ is the ratio of the flapping cycle during which both the leader and the follower are in their downstroke configurations (and so on for the remaining three weights). The total wake-induced time-averaged force on the follower is thus
\begin{equation}\label{eqn:total_induced_force}
    \vec{F}^I = \beta_{\rm{dd}}\vec{F}_{\rm{dd}} + \beta_{\rm{du}}\vec{F}_{\rm{du}} + \beta_{\rm{ud}}\vec{F}_{\rm{ud}} + \beta_{\rm{uu}}\vec{F}_{\rm{uu}}.
\end{equation}
This may be substituted into Eq. (\ref{total_follower_force}) to model the total time-averaged force on the follower.

We solve for $\Gamma$ by imposing $F_z^L=Mg=W$, i.e., the leader must support its own weight via the vertical component of its generated force. From Eqs. (\ref{leader_vert_force}-\ref{leader_vert_imp}), this yields
\begin{equation}\label{eq:gamma_sol}
\left(\frac{\pi}{4} + R\right)\Gamma = \frac{WT}{\rho b \lambda} = \frac{W}{\rho b U},
\end{equation}
which admits a unique solution for $\Gamma$. Then, with $\Gamma$ defined, the requirement $F_x^L = {I_x^L}/{T^L} = D$ determines a constant value for the drag $D$ that each bird must oppose. The circulation is thus directly proportional to the bird's weight and inversely proportional to the air density, wingspan, and flight speed. 

In Eq. \ref{eq:gamma_sol}, the prefactor $(\pi/4 + R)$ naturally separates the contributions of the downstroke ($\pi/4$) and the upstroke ($R$). The retraction ratio, $R$, only appears in the model in the product $R\Gamma$, and thus provides an implicit control on the effective upstroke circulation. For example, $R=0$ corresponds to a passive upstroke, effectively eliminating upstroke circulation.

\section*{Optimization}

Using this wake-vortex force model for a leader-follower pair of birds, we conduct a constrained numerical optimization to determine the follower's most advantageous three-dimensional reference coordinates $(x_F,y_F,z_F)$, tip-to-tip flapping amplitude $h_F$, upstroke flexion ratio $R_F$, and temporal flapping phase offset parameter $\kappa_F$. As such, the optimization is conducted over a 6-dimensional parameter space.  The position and kinematics of the leader $(x_L,y_L,z_L, h_L,R_L,\kappa_L)$ are prescribed, as are the environmental conditions -- $U$, $\rho$, $g$ -- and species conditions -- the bird mass $M$, wing half-span $b$, the wing chord $c$, and the wingbeat period $T$. The implications of these assumptions are discussed in later sections.

There are several possible choices for a cost function to minimize, subject to the  constraints that the follower must achieve weight support and match the leader's speed by satisfying thrust-drag balance. Here, we explore two possible choices of cost function: one encoding force, and the other, power. To minimize the total time-averaged force required by the follower, the cost function, $C$ is given by:
\begin{equation}\label{objective_fn}
    C = \sqrt{I_x^F(h_F,R_F)^2 + I_z^F(h_F,R_F)^2}
\end{equation}
where $I_x^F$ and $I_z^F$ are respectively given by Eqs. (\ref{leader_vert_imp}) and (\ref{leader_hor_force}).

A second choice of cost function encodes the total mechanical power expenditure of the follower. To quantify this, we define the induced power (the power to generate lift to counter gravity), adapted directly from Pennycuick \cite{pennycuick2008modelling}, as
\begin{equation}\label{induced_power}
    P_{\rm ind} = \frac{2kI_z^F(h_F,R_F)^2}{(2b)^2 \pi\rho UT^2}
\end{equation}
where $k=1.2$ is the induced power factor \cite{pennycuick2008modelling}. Similarly, we define the profile power (the power to generate thrust to counter drag associated with the wings and wing motion) as
\begin{equation}\label{profile_power}
    P_{\rm pro} = \frac{I_x^F(h_F,R_F)U}{T}.
\end{equation}
Finally, we define the parasite power (the power required to overcome the body drag), also adapted directly from \cite{pennycuick2008modelling}, as
\begin{equation}\label{parasite_power}
    P_{\rm par} = \frac{\rho U^3 S_b C_{D_b}}{2}
\end{equation}
where $S_b=0.00813 M^{0.666}$ and $C_{D_b}=1$. Thus, the objective function to be minimized is
\begin{equation}
    C = P_{\rm ind}(h_F,R_F) + P_{\rm pro}(h_F,R_F) + P_{\rm par}.
\end{equation}

For either choice of cost function, the minimization is subject to two constraints, namely
\begin{align}
    \frac{I_x^F(h_F,R_F)}{T} + F_x^I(x_F,y_F,z_F,h_F,R_F,\kappa) &= D, \\
    \frac{I_z^F(h_F,R_F)}{T} + F_z^I(x_F,y_F,z_F,h_F,R_F,\kappa) &= W.
\end{align}
That is, the sum of the follower's self-generated force from flapping ($I_{x,z}^F/T$) and its wake-induced force from the leader ($F_{x,z}^I$) must balance its drag $D$ and weight $W$. Thus, although both the leader and follower must attain force balance against identical drag $D$ and weight $W$ in order to sustain flight at constant speed $U$, the follower may take advantage of the wake of its leader and thereby reduce the force that it must generate on its own by flapping.

The following search bounds and initial values are applied for the optimization:
\begin{enumerate}
    \item The follower is behind its leader in the streamwise direction (symmetry): $x_{F} \leq x_L$. The supplied initial value is $x_F^0 = x_L$.
    \item The follower is to the right of its leader in the spanwise direction (symmetry): $y_F \geq y_L$. The supplied initial value is $y_F^0 = y_L$.
    \item The follower's vertical position is relatively near the leader's \cite{portugal2014upwash}: $z_L-h_L \leq z_F \leq z_L+h_L$. The supplied initial value is $z_F^0 = z_L$.
    \item The follower's flapping amplitude is within 75\% of that of a lone bird: $0.25h_L\leq h_F \leq h_L$. Note that the upper bound is set to $h_L$, as the objective function $C$ decreases monotonically with $h_F$, i.e., $h_F>h_L$ cannot be optimal. To our knowledge, no data exists to characterize birds' flapping amplitude during formation flight. However, birds do vary their flapping amplitude on the order of 20-30\% in response to aerodynamic and energetic contexts \cite{krishnan2022role}, and so we choose a generous lower bound accordingly. The supplied initial value is $h_F^0 = h_L$. 
    \item The follower's upstroke flexion ratio is within reasonable physical bounds \cite{tobalske1996flight}: $0.3 \leq R_F \leq 0.5$. The supplied initial value is $R_F^0 = R_L$.
    \item No search bounds are applied to the follower's phase parameter: $\kappa_F\in[0,1)$. The supplied initial value is $\kappa_F=0.5$.
\end{enumerate}

The optimization is performed using the \verb|fmincon| solver in MATLAB, and we find that the optimization results are generally robust to changes in search bounds and initial values, suggesting that the attained minimum is global. Details of the numerical implementation are provided in Materials and Methods.

\section*{Results}

\begin{table*}[t!]
\centering
\begin{tabular}{llll}
Quantity & Leader & Follower (our results) & Follower (Portugal et al. \cite{portugal2014upwash}) \\
\midrule
Streamwise position (m) & $x_L=0$ & $x_F=1.83$ & $x_F\approx 1.2$ \\
Spanwise position (m) & $y_L=0$ & $y_F=0.882$ & $y_F\approx 0.904$ \\
Vertical position (m) & $z_L=h_L/2=0.520$ & $z_F=0.457=0.88z_L$ & $z_F\approx z_L$ \\
~&~&~\\
Relative upstroke span & $R_L=0.4$ & $R_F = 0.347=0.87 R_L$ & NR  \\
Tip-to-tip amplitude (m) & $h_L=1.04$ & $h_F = 0.751=0.72h_L$ & NR  \\
Temporal phase offset ratio & $\kappa_L=0$ & $\kappa_F=0.500$ & NR \\
~ & ~& ~  \\
Self-generated lift (N) & $I_z^L/T = W=12.74$ & $I_z^F/T=12.17=0.96W$ & NR\\
Self-generated thrust (N) & $I_x^L/T=D=2.30$ & $I_x^F/T=1.89=0.83D$ & NR \\
\bottomrule
\end{tabular}
\caption{Summary of numerical results from optimization, with direct comparison against the experimental live-ibis results of Portugal et al. \cite{portugal2014upwash} as reported directly in the work. The notation ``NR'' means that, to our knowledge, the quantity is not reported. Note that Portugal et al. do not report a $z_F$ value for follower birds, noting only that ``birds were flying close
to the same horizontal plane" during flight \cite{portugal2014upwash}. We denote this finding here as $z_F\approx z_L$.}\label{table:table1}
\end{table*}

Here we give illustrative results of our model and optimization procedure for a specific migratory bird species: the northern bald ibis (\textit{Geronticus eremita}). The following parameters are taken directly from the experimental observations of Portugal et al. \cite{portugal2014upwash}, who reported flocking behaviors for the same species (all units kg$\cdot$m$\cdot$s): gravitational acceleration $g=9.8$, bird mass $M=1.3$, air density $\rho=1.29$, semispan length $b=0.6$, average flight speed $U=15$, and flapping period $T=1/4$. We assign also a dimensionless upstroke flexion ratio of $R_L=0.4$ \cite{tobalske1996flight}. Finally, by inspection of figures and movies in \cite{portugal2014upwash,DisneyWhiteIbis2025}, we assign the chord length $c=b/4$ and the flapping amplitude $h_L=2b\sin(\pi/3)$. With these values defined and applying Eq. (\ref{eq:gamma_sol}), this yields a weight $W=12.7$ N and drag $D=2.3$ N for each ibis to oppose during flight.

\begin{figure}[t!]
\centering
\includegraphics[width=\linewidth]{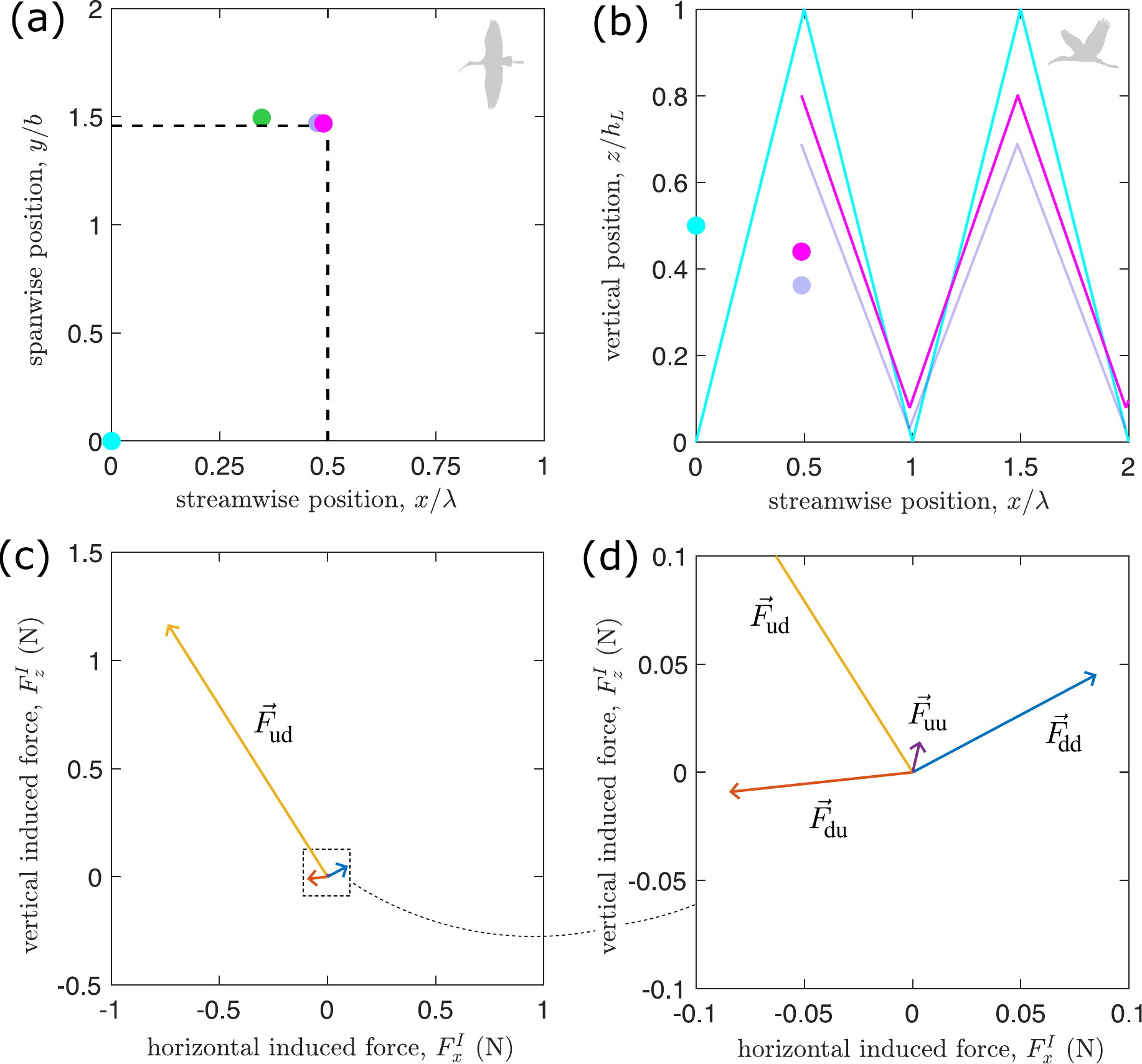}
\caption{Results of optimization procedure. (a) Optimal position of the follower (force minimization, magenta; power minimization, purple) relative to its leader (cyan) in the dimensionless $(x/\lambda, y/b)$ plane. The average experimental observation of Portugal et al. \cite{portugal2014upwash} is shown by a green dot. The dashed vertical line has equation $x/\lambda=0.5$, and the dashed horizontal line has equation $y/b=\pi/4 + (1+R_F)/2$. Bird silhouettes are meant to orient the reader and are not to scale. (b) Optimal position in the dimensionless $(x/\lambda, z/h_L)$ plane. The extrapolated wakes of the leader (cyan) and follower (force minimization, magenta; power minimization, purple) align closely, demonstrating leader-follower wingtip path coherence in the optimal solution. (c,d) Wake-induced force components, computed at the optimal solution, which are used in Eq. (\ref{eqn:total_induced_force}) to determine the total force on the follower. Panel (d) is an inset of panel (c). }
\label{fig:result_v}
\end{figure}

The results of the optimization procedure are summarized by Table \ref{table:table1} and Fig. \ref{fig:result_v}, which together demonstrate that our predictions broadly align with the live-ibis experimental observations of Portugal et al. \cite{portugal2014upwash}.

Of major significance is that the two choices of cost function -- one which encodes the follower's total force expenditure, and the other its mechanical power expenditure -- produce almost identical results. This similarity is evident in Fig. \ref{fig:result_v}(a,b), which overlays results from directly minimizing power expenditure against those from minimizing force production. Whether this similarity persists for other avian species (i.e., for other choices of parameters in the model) presents a potentially fruitful line of inquiry that may be explored using this framework. Exploring the effect of other cost functions (which involves changing just one line of code in the present framework) would also be highly valuable.

\subsection*{Wingtip path coherence}

Figure \ref{fig:result_v}(a) shows that the optimal streamwise location of the follower, which minimizes total force production, is $x_F \approx \lambda/2$, and the optimal temporal phase offset ratio is $\kappa_F=0.5$. That is, the follower flies slightly less than half a stroke wavelength behind the leader, and it flaps its wings in temporal antiphase with respect to the leader. Taken together, these results indicate that the follower synchronizes its wingbeat with the phase of the leader's wake by selecting both its position and flapping phase so that its wingtips remain aligned with coherent structures shed by the leader (Fig. \ref{fig:result_v}(b)). This behavior is known as wingtip path coherence, and it has been proposed as a mechanism for sustained upwash capture, with support from both computational and live-bird studies \cite{portugal2014upwash, willis2007computational,friman2024pays}.

With our model, the underlying rationale for the optimality of this wingtip path coherence is revealed. Within the framework of Eqs. (\ref{eqn:beta_uniform}-\ref{eqn:total_induced_force}), setting $\kappa_F = 0.5$ yields $\beta_{\rm{dd}} = \beta_{\rm{uu}} = 0$ and $\beta_{\rm{du}} = \beta_{\rm{ud}} = 0.5$. Consequently, the force contributions $\vec{F}_{\rm{uu}}$ and $\vec{F}_{\rm{dd}}$ receive zero weight in the total induced force. As shown in Fig. \ref{fig:result_v}(c,d), these two components are precisely those with a positive downstream (drag-producing) contribution. Therefore, the optimal wingtip path coherence configuration effectively eliminates any additional drag which could arise from wake interaction with the leader. This model outcome supports the hypothesis that real birds may exploit the same synchronization strategy during unsteady formation flight.

Finally, as shown in Fig. \ref{fig:result_v}(a), our prediction (magenta and purple dots) of the streamwise position of the follower $x_F$ is moderately larger than that which is reported experimentally (green dot). This discrepancy may be due to our model's neglecting of biological drives -- such as an innate desire to remain close to other flock members, which may be triggered by visual cues \cite{fernandez2004visual} -- or due to complexities in the true vortical wake not captured by this minimal model. Regardless, the prediction (Table \ref{table:table1}, $x_F=1.8$ m) is reasonably close to the experimentally-reported value of $x_F=1.2$ m, and indeed, this experimental value itself displays high variation during actual ibis flight \cite{portugal2014upwash}. 

\subsection*{Global structure of wake-induced forces}

\begin{figure}[t!]
\centering
\includegraphics[width=\linewidth]{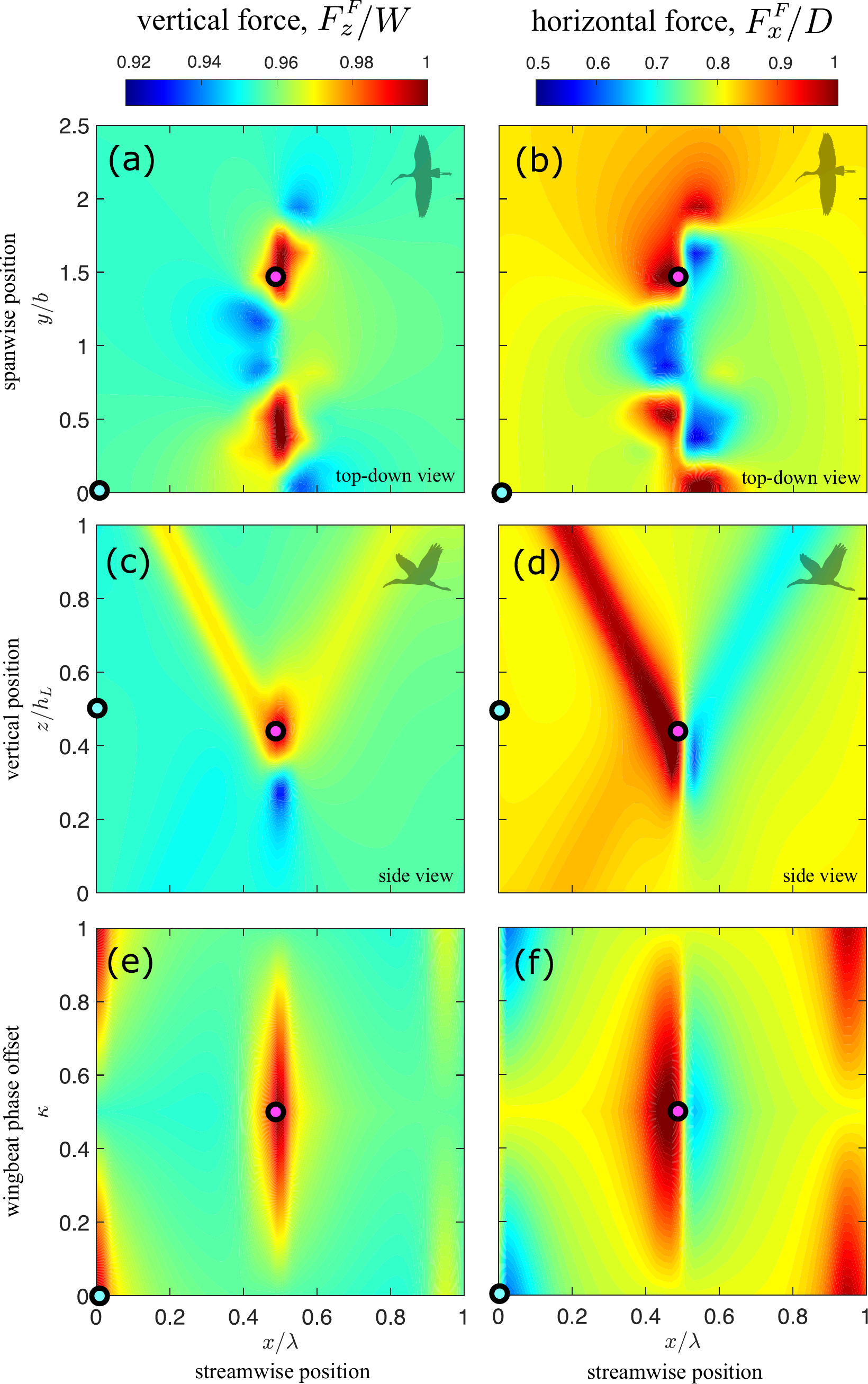}
\caption{Planar sections of normalized vertical and horizontal net forces $F_z^F/W$ (a,c,e) and $F_x^F/D$ (b,d,f) on the follower. The follower's optimal dimensionless location in state space is shown by a magenta dot in each section. This location precisely satisfies $F_z^F/W = F_x^F/D=1$. The leader's location is shown by a cyan dot at $(x_L,y_L,z_L)$. Force balance is more closely satisfied in regions of red, whereas the follower produces insufficient force in non-red regions. The variation of $F_x^F$ is much larger than that of $F_z^F$, even dropping as low as 50\% of the total thrust requirement -- while the vertical force only drops as low as about 90\% of the total lift requirement. Bird silhouettes are meant to orient the reader and are not to scale.}
\label{fig:force_surfaces}
\end{figure}

A central strength of the present model is that it provides the full net force on the follower, $\vec{F}^F$, as an explicit function over a six-dimensional state space: spatial position $(x,y,z)$ and flapping parameters $(h,R,\kappa)$. Fig. \ref{fig:force_surfaces} shows representative two-dimensional slices of this force field. The horizontal and vertical components are respectively normalized by drag $D$ and weight $W$. In these plots, $(z,\kappa,h,R)$ are fixed at their optimal values $(z_F,\kappa_F,h_F,R_F)$, and the surfaces represent total forces (self-generated plus wake-induced) as defined in Eq. \ref{total_follower_force}.

Panels \ref{fig:force_surfaces}(a,b) show the force landscape in the normalized $(x/\lambda,y/b)$ plane of streamwise versus spanwise position. The field is highly structured, with alternating regions where the wake interaction is beneficial or detrimental. Crucially, regions that provide a thrust benefit do not generally coincide with those that provide a lift benefit. For example, near $y=0$, the wake induces a forward force but a downward force. The optimal location $(x_F,y_F)$ (magenta marker) lies where favorable vertical and horizontal contributions overlap most strongly. This demonstrates that the common ``upwash/downwash'' heuristic is incomplete for unsteady flapping: a region of upwash is not necessarily advantageous in both lift and thrust, in contrast to the simpler fixed-wing picture.

Panels \ref{fig:force_surfaces}(c,d) show the landscape in the normalized $(x/\lambda,z/h_L)$ plane of streamwise versus vertical position. At $x=\lambda/2$, the vertical force changes sign across the leader's height, with upward force above and downward force below. At the same time, the horizontal force changes sign in the streamwise direction: just upstream of $x=\lambda/2$, the interaction is thrust-producing, while just downstream it is drag-producing. This explains why the streamwise optimum lies slightly upstream of $\lambda/2$; at exactly $\lambda/2$, the horizontal wake-induced force transitions from advantageous to disadvantageous.

Panels \ref{fig:force_surfaces}(e,f) present the normalized $(x/\lambda,\kappa)$ plane of streamwise position versus wingbeat phase and demonstrate the advantageous role of wingtip path coherence. Favorable force regions align along combinations of streamwise position and phase that keep the follower's wingtips in phase with the leader's wake structures. The optimum at $\kappa=0.5$ and slightly upstream of $x=\lambda/2$ lies on this alignment.

Finally, the strongly beneficial regions (dark red) occupy only a small fraction of the domain. This reflects the optimization objective (Eq. (\ref{objective_fn})), which drives the follower to minimize its own lift and thrust production. Away from favorable wake interactions, the follower's reduced amplitude and upstroke ratio are insufficient to maintain force balance; only in specific regions of the wake field does the induced force supply the deficit required for steady flight.

\section*{Analysis}

We now analyze the predictions of our model and optimization procedure, and consider their implications in the context of general avian formation flight.

\subsection*{Optimal spanwise spacing}

As shown in Table \ref{table:table1}, our model reports an optimal spanwise position $y_F=0.877$ which is just slightly lower than that reported experimentally by \cite{portugal2014upwash}, $y_F=0.904$. This slight discrepancy may be explained by noting that real birds may accept being slightly further away from the maximally beneficial region to minimize the risk of straying into a zone where savings are negative \cite{cutts1994energy}.

We seek an approximate analytical expression for the optimal spanwise offset predicted by our model. The guiding physical assumption is that performance is maximized when the follower's wingtips remain as close as possible to the leader's upwash region. Accordingly, the optimal $y_F$ should minimize, in a cycle-averaged sense, the spanwise distance between a follower wingtip and the relevant wake vortex.

\begin{figure}[t!]
\centering
\includegraphics[width=\linewidth]{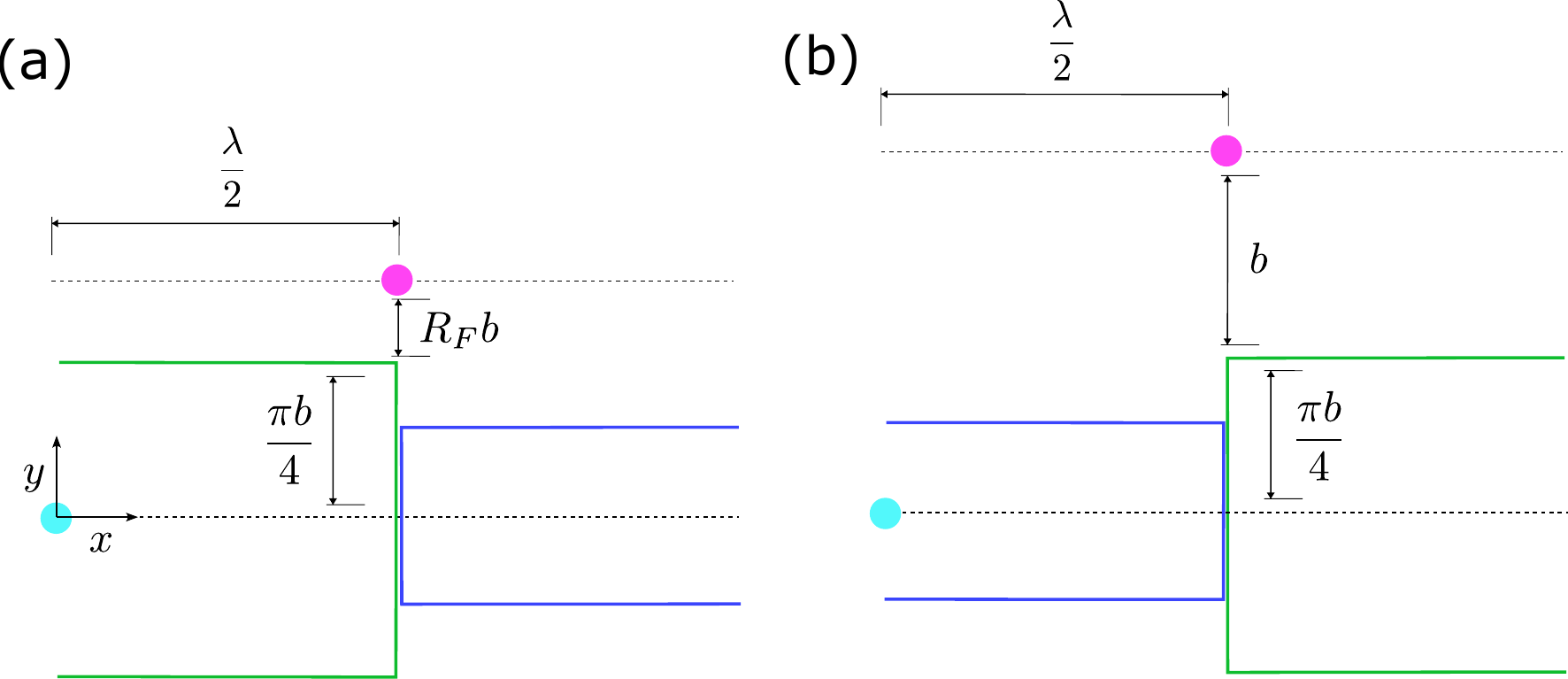}
\caption{Schematic of the follower's (magenta) optimal spanwise positioning relative to the leader (cyan) in the down-up (a) and up-down (b) model configurations. This optimal position is such that the follower's wingtips lie just outside of the nearest streamwise vortex filament. Downstroke and upstroke wake vortex filaments are shown respectively in green and blue. The cycle-averaged optimal distance is then the average of these two distances. (a) Down-up configuration. The optimal wingtip-to-vortex spanwise distance is $y = \pi b/4 + R_F b$. (b) Up-down configuration. The optimal wingtip-to-vortex spanwise distance is $y=\pi b/4 + b$.}
\label{fig:spanwise}
\end{figure}

As described in the previous subsection, the optimal phase relation $\kappa_F=0.5$ restricts the interaction to the up-down and down-up configurations (Fig. \ref{fig:wake_structure}(b,c)), each occupying half of the wingbeat period. The streamwise optimum $x_F\approx \lambda/2$ places the follower adjacent to the leader's downstroke-generated streamwise vortex segments, located at $y=\pi b/4$ throughout the cycle (Fig. \ref{fig:spanwise}). Over one cycle, the follower performs an upstroke for half the period and a downstroke for the other half. The spanwise distance from the follower's centerline to its wingtip is $R_Fb$ during the its upstroke and $b$ during its downstroke. Thus, the cycle-averaged optimal spanwise position is obtained by averaging the corresponding optimal wingtip-to-vortex distances:
\begin{equation}\label{optimal_spanwise_dist}
    y_F = \frac{1}{2}\left(\frac{\pi b}{4}+b + \frac{\pi b}{4}+R_Fb\right).
\end{equation}
This analytical form agrees with the numerical optimum to within 1\%. The small discrepancy is physically consistent: the wingtip must lie slightly inside the upwash region rather than exactly at the geometric meeting point implied by the formula, which explains why the computed value $y_F=0.886$ is marginally larger than the analytic value $y_F=0.877$.

This reasoning suggests a more general expression. Let $w_L(t,x_F)$ denote the spanwise distance from the leader's centerline to the relevant wake vortex at the follower's streamwise position $x_F$ at time $t\in[0,T)$. In the present configuration, $w_L(t)=\pi b / 4$. Then let $r_F(t)$ be the spanwise distance from the follower's centerline to its wingtip. In our case, this is
\begin{equation}
    r_F(t) = \begin{cases}
        b & 0\leq t < T/2 \\
        bR_F & T/2 \leq t < T.
    \end{cases}
\end{equation}
A natural generalization for the optimal spanwise offset is then
\begin{equation}
    y_F = \frac{1}{T}\int_0^T (w_L(t) + r_F(t)) dt.
\end{equation}
that is, the time average over one cycle of the sum of the wake-vortex location and the instantaneous wingtip offset. Substituting the present forms of $w_L$ and $r_F$ into the above yield Eq. (\ref{optimal_spanwise_dist}). Future work might investigate the accuracy of this generalized form based on experimental measurements of $w_L(t)$ and $r_F(t)$ during live-bird flight.

\subsection*{Power reduction via formation flight}

As reported in Table \ref{table:table1}, the follower bird is only required to generate 96\% of the lift and 83\% of the thrust required of a leader bird of the same size and mass. Writing the total mechanical power to fly as $P = P_{\rm ind}+P_{\rm pro} + P_{\rm par}$ (Eqs. (\ref{induced_power}-\ref{parasite_power})), this gives $P_L = 60$ W for the leader and $P_F = 53$ W for the follower. The wake interaction therefore yields an 11\% reduction in total power required by the follower. This magnitude is consistent with field and laboratory estimates for comparably large birds (e.g., geese and pelicans) flying in formation, which generally report energetic savings on the order of 10-15\% \cite{voelkl2015matching, cutts1994energy, hainsworth1988induced}.

A key advantage of the present framework is that it resolves the total power saving into aerodynamic components. The follower experiences a 9\% reduction in induced power and an 18\% reduction in profile power (the parasite power is constant for both birds). Such a decomposition is effectively inaccessible in experiments. Empirical studies, e.g., \cite{weimerskirch2001energy}, rely on global physiological or kinematic proxies such as heart rate or wingbeat frequency, which reflect total metabolic or mechanical output but cannot distinguish whether the savings arise primarily from reduced lift production (induced power) or reduced drag and wing work (profile power). This limitation helps explain the variation and ambiguity in the literature, where formation benefits are variously described in terms of ``extra lift'' \cite{voelkl2015matching}, ``reduced induced drag'' \cite{hummel1983aerodynamic,hainsworth1988induced}, or broadly ``upwash exploitation'' \cite{portugal2014upwash}, often without mechanistic specificity.

The present results clarify this issue: the follower benefits through both channels, but the dominant contribution in this regime is a reduction in profile power rather than induced power. In other words, the wake interaction does not merely offload weight support; it more substantially reduces the aerodynamic resistance and work associated with flapping.

This distinction leads directly to the kinematic mechanism. Statements that followers ``exploit upwash'' describe the flow-level cause but not the bird-level realization. In the present model, the aerodynamic assistance allows the follower to reduce force production at the wing, which is implemented primarily through a reduction in flapping amplitude $h_F$ and, secondarily, through a reduction in the upstroke flexion ratio $R_F$. The amplitude reduction is larger (28\%, Table \ref{table:table1}) than the reduction in $R_F$ (13\%), indicating that decreased stroke amplitude is the principal kinematic pathway by which aerodynamic benefits translate into mechanical power savings. Thus, the model connects wake aerodynamics to specific, quantifiable changes in wing motion that together lower both induced and profile power.

\subsection*{Multi-bird flocks}

A natural extension of this model involves iteratively adding more birds into the flock to solve for the optimal position and flapping kinematics of a multi-tiered formation rather than just a single leader-follower pair. This is achieved easily within the presented framework by solving the optimization problem $N-1$ times (i.e., considering $N-1$ successive leader-follower pairs) for a flock of $N$ birds, with the follower at iteration $i$ becoming the leader at iteration $i+1$. 

\begin{figure}[t!]
\centering
\includegraphics[width=\linewidth]{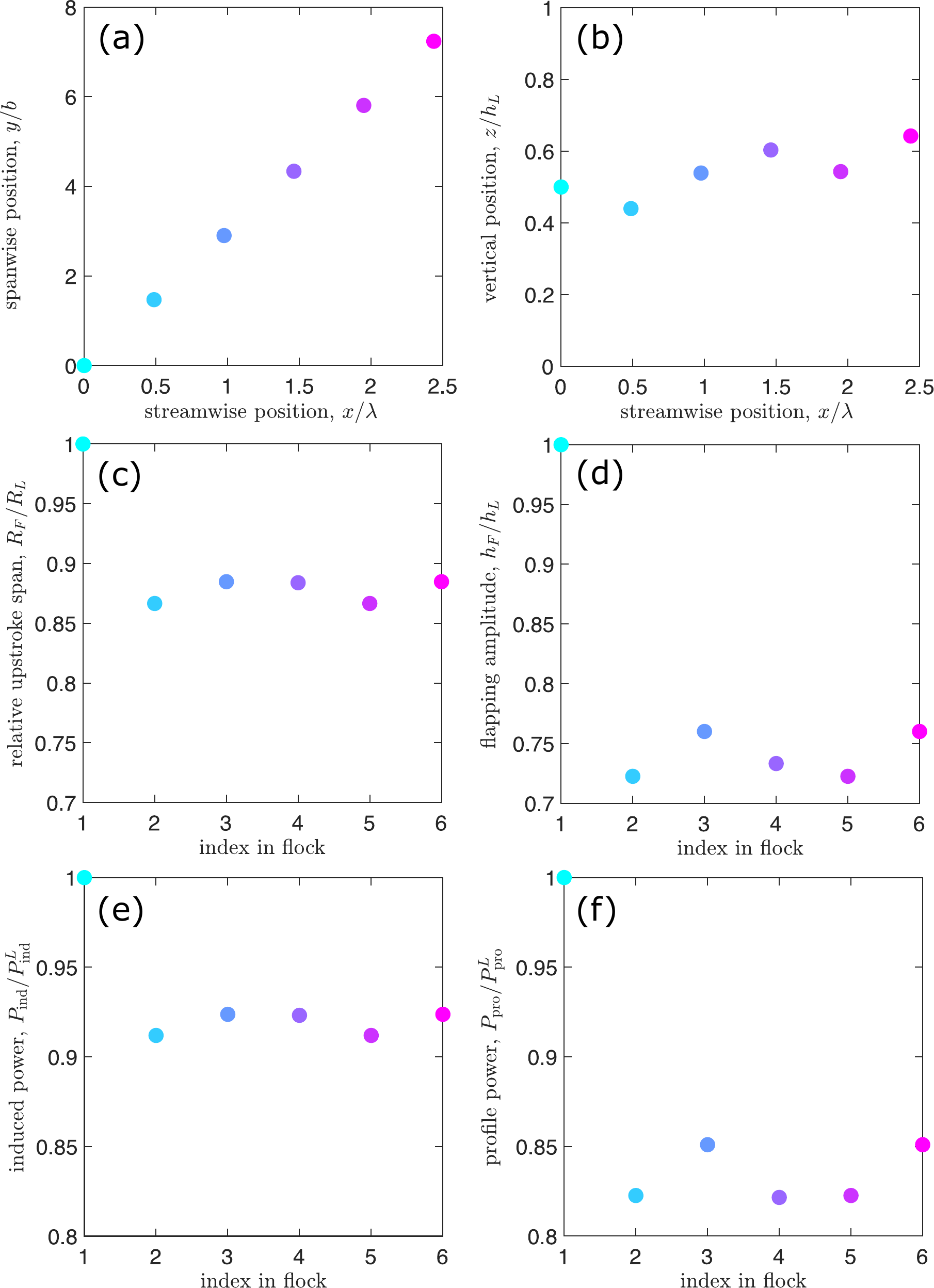}
\caption{Six-bird formation of northern bald ibises, produced by repeated optimization. (a,b) Optimal positioning in the $(x,y)$ (a) and $(x,z)$ (b) planes. (c,d) Kinematic parameters $R_F$ (c) and $h_F$ (d) of each successive flock member, normalized against that of the flock leader ($i=1$). (e,f) Induced (a) and profile (b) power of each successive flock member, normalized against that of the flock leader.}
\label{fig:multi_tier}
\end{figure}

Fig. \ref{fig:multi_tier} presents illustrative results for a formation of six birds. Evidently, the streamwise-spanwise planar spacing is highly regular (Fig. \ref{fig:multi_tier}(a)). Moreover, the optimal leader-follower temporal phase offset $\kappa$ persists throughout the flock, with each follower flapping in precise temporal antiphase with respect to their leader: $\kappa_{i+1} = \mod(\kappa_i + 0.5,1)$.

While aerodynamic benefits are maximized for the second bird in the flock $(i=2)$, downstream birds continue to accrue substantial gains. However, these benefits do not decay monotonically with position in the formation; instead, they rapidly saturate with low variation (Fig. \ref{fig:multi_tier}(e,f)). This is readily explained by the earlier analysis. Each follower can extract a comparable energetic advantage from its immediate leader because the dominant mechanism is interaction with the leader’s downstroke wake, which provides both lift and thrust augmentation (Fig. \ref{fig:result_v}(c,d) and Fig. \ref{fig:spanwise}). Each follower adjusts its own kinematics -- its upstroke span ratio $R$ and flapping amplitude $h$ -- to exploit this flow (Fig. \ref{fig:multi_tier}(c,d)). Since $R$ affects only the upstroke, it has little impact on the ability of each follower to benefit from its leader’s downstroke wake. The parameter $h$ does influence the inclination of the wake, but the key geometric requirement is that the follower’s wingtips remain aligned with the edge of the downstroke wake (Fig. \ref{fig:spanwise}). This alignment can be achieved primarily through vertical positioning $z$, which is a free parameter. As a result, each bird in the formation can maintain similar energetic benefits by optimally positioning itself within the downstroke wake of its immediate predecessor. As is evident in Fig. \ref{fig:multi_tier}(b), this vertical positioning displays some variation, which illustrates this effect.

\section*{Discussion}

This work develops a time-averaged theoretical framework for flapping formation flight based on a single leader-follower pair of birds. Using an impulse-based formulation as in \cite{spedding2003family} for the forces generated during the upstroke and downstroke, we construct a simplified three-dimensional representation of the unsteady wake shed by the leader and couple it to a follower whose spatial position and flapping parameters are determined through optimization. The objective is to retain the essential unsteady aerodynamic interactions while avoiding the complexity of fully resolved simulations, e.g., \cite{beaumont2022modeling,beaumont2024aerodynamic,willis2007computational}. The result is a hybrid description in which the flapping cycle and wake evolution are represented as a sequence of quasi-static vortex configurations, extending classical impulse-based wake models for a single bird to a formation setting.

Despite its simplicity, the model reproduces the canonical V-formation geometry of migratory birds, aligning quantitatively with experimental live-bird observations \cite{portugal2014upwash}. The optimality of this formation is explained via direct interrogation of the global structure of the wake-induced forces on a bird which interacts with its leader's wake. In doing so, the model yields physically interpretable mechanisms for wake exploitation. It supports the hypothesis of wingtip path coherence \cite{portugal2014upwash,willis2007computational}, whereby the follower synchronizes both position and phase with the leader's wake structures, and it provides a force-field view showing that regions favorable for thrust and lift do not necessarily coincide. Optimal positioning therefore emerges as a multi-objective compromise in a six-dimensional state space, rather than a simple rule of ``sitting in upwash'' as is generally noted in existing studies of formation flight \cite{voelkl2015matching,hummel1983aerodynamic,hainsworth1988induced,portugal2014upwash,friman2024pays}.

The framework also enables a force- and power-resolved interpretation of energetic savings that is currently inaccessible experimentally. While empirical studies can infer total energy reduction through physiological or kinematic proxies such as heart rate \cite{weimerskirch2001energy}, they cannot reliably partition savings between induced and profile power. Here, the predicted reduction in total power of 11\% is consistent with reported formation flight benefits of 10-15\% \cite{hainsworth1988induced,cutts1994energy,voelkl2015matching}, but the mechanism is clarified: the follower experiences reductions in both induced and profile power, with profile power reduction playing the dominant role (17\% reduction) rather than induced power (8\% reduction).

Crucially, the model links aerodynamic assistance to specific kinematic adjustments, providing a concrete realization of how ``upwash exploitation'' translates into altered force production at the wing. Classical fixed-wing analyses \cite{lissaman1970formation, hummel1983aerodynamic, hainsworth1988induced} show that intercepted upwash produces an immediate forward rotation of the follower's aerodynamic force, yielding a direct reduction in induced drag or an effective thrust benefit. Experimental studies and high-fidelity CFD likewise identify favorable wake interactions, but generally infer their benefit only in aggregate or describe them qualitatively as enhanced lift support \cite{portugal2014upwash,voelkl2015matching, willis2007computational}. The present flapping formulation shows that both effects are operative: leader-generated upwash supplies a drag-relieving benefit as predicted by fixed-wing theory, while also furnishing a vertical support contribution through unsteady wake-wing interactions. Because the follower requires less self-generated horizontal and vertical force, it re-satisfies force balance through modified stroke kinematics, principally via reduced flapping amplitude and secondarily via reduced upstroke flexion. The diminished flapping effort lowers both induced and profile power, but the larger reduction occurs in profile power, indicating that the realized energetic savings arise primarily through mitigation of work-intensive flapping excursions. The energetic benefit of formation flight should therefore be interpreted not as a single mechanism, but as a coupled aerodynamic and kinematic redistribution of effort whose relative components can now be quantified. This connection between wake aerodynamics, force balance, and stroke-level kinematics addresses a gap left by prior studies, which typically identify the flow mechanism but not its mechanical implementation by the bird. It may be that this kinematic mechanism is universally realized for birds which participate in group flight, or that this is a species-specific phenomenon linked to some particular aspects of wing morphology. Detailed kinematic data on birds flying alone versus within a flock are necessary.

It is important to note that in the present model, we consider two choices of objective function in the optimization step: first, the total aerodynamic force, and second, the total power expenditure. There is no consensus in the flight biomechanics literature on a single objective that birds optimize during sustained group flight. First, non-aerodynamic aspects such as social dynamics or predator avoidance likely factor into avian decision-making during formation flight \cite{voelkl2015matching, fernandez2004visual, salahshour2025allocentric}. These aspects aside, studies variously posit minimization of power (instantaneous energetic expenditure) \cite{bishop2024flying}, cost of transport (energy per unit distance) \cite{pennycuick1969mechanics}, maximization of endurance/range (time or distance achievable with finite energy stores) \cite{sachs2012flying}, or aerodynamic efficiency (producing required forces with minimal dissipation) \cite{johansson2024aerodynamic}. These objectives are not equivalent: they weigh different aspects of the same underlying physics, namely the relationship between wing kinematics, force production, and energetic cost. In particular, minimizing the magnitude of self-generated force isolates how effectively a bird can offload aerodynamic load onto the surrounding flow, whereas minimizing power quantifies the energetic expense of producing that load. These two objectives -- force minimization and power minimization -- are related but not equivalent, as power scales nonlinearly with vertical force but linearly with horizontal force \cite{pennycuick2008modelling}. However, we demonstrate that the results from optimizing with respect to both of these cost functions align closely.

A third reasonable choice of cost function might be the energetic cost of transport during long-haul flight. This cost of transport is defined as $E = P/U$ \cite{hedenstrom1993migration}, where $P$ is the total mechanical power and $U$ is the flight speed. In minimizing the cost of transport, future work might treat $U$ itself as an optimization variable rather than a fixed constant. Such a reformulation could refine the over-prediction of streamwise position observed in the current model, while also directly predicting the optimal cruising speed for formation flight. Encouragingly, preliminary explorations within this framework suggest that this predicted optimal group speed -- which minimizes the total mechanical power of the follower -- is lower than that of an individual bird, which is consistent with existing theoretical predictions based on simplified fixed-wing models \cite{lissaman1970formation,kshatriya1992theoretical}.

Moreover, within the framework developed here, we assume that $P$ and $U$ are constant. But, in nature, migratory birds adjust many aspects of their flight conditions, including speed and altitude (which changes air density) during flight.  A significant variable during long flights is weight: birds may lose as much as 20-40\% of their body weight during migration \cite{nisbet1963weight, straker2017rapid}, which reduces lift requirements and can shift both the optimal speed and the optimal formation configuration over time. The true cost of transport is therefore $E = \int_t P(t)dt ~/ \int_t U(t)dt$, where both $P(t)$ and $U(t)$ depend on instantaneous body weight and local atmospheric conditions \cite{pennycuick1969mechanics}. Future work might therefore refine or augment the present framework to account for this time dependence during long-haul flight and for the likelihood that birds prioritize lift-based versus thrust-based assistance from wake interactions differently as migration progresses.

Future work might also systematically explore the effects of parametric variation and the use of different cost functions (total aerodynamic force, power expenditure, cost of transport, or others) on the formation of multi-bird flocks, a single example of which is demonstrated in Fig. \ref{fig:multi_tier}. Investigations of this nature might reveal, for a given set of parameters and a given cost function to optimize, whether energetic benefits decay or hold stable throughout a flock, whether flocks grow disorganized or remain in a regular formation, whether the optimal speed varies based on flock size, and so on.

In summary, the present model isolates the aerodynamic component of the formation-flight optimization problem and therefore identifies an energetically favorable configuration in the absence of additional ecological or behavioral constraints. The model's limitations follow directly from its intended level of abstraction. Some simplifications, such as the constant circulation $\Gamma$ during the upstroke and downstroke, are simply due to a lack of empirical data. However, with further data, this assumption could be easily modified -- and, as the upstroke circulation is scaled by the upstroke contraction factor $R$ in this model (Eq. \ref{eq:gamma_sol}), the upstroke kinematics in a sense modulate this fixed-circulation assumption.

Other simplifications are made for tractability. For example, changes in profile drag associated with modified kinematics are not modeled explicitly, and the wake is represented by idealized vortex structures sampled at discrete stroke phases. An immediate next-generation iteration of this model might introduce intermediate states at the mid-stroke, so that the leader and follower can each access three available states (up, down, and middle), leading to nine possible leader-follower configurations. In general, allowing the leader and follower to each access $N$ states leads to $N^2$ leader-follower configurations. Fidelity of the model to true flapping should increase with $N$, but there is a major tradeoff between fidelity and computational tractability. It is important to note that traditional fixed-wing models involve $N=1$, and this model, which involves $N=2$, marks a significant step towards accounting for real unsteadiness. It is difficult to assess a priori the further accuracy accrued by increasing $N$ further; this lies outside the scope of the present work.

The model also neglects time dependence of birds' weight, body aerodynamics, viscous wake decay, detailed vortex deformation, wing flexibility, feather-scale effects, and non-aerodynamic biological drivers such as social spacing preferences, vision, leadership rotation, and predator avoidance \cite{fernandez2004visual, voelkl2015matching, salahshour2025allocentric, caraco1980avian}. These omitted constraints would not eliminate the aerodynamic optimum, but would act as competing objectives that may displace the realized \textit{in vivo} formation away from the purely energetic optimum \cite{cutts1994energy}. For example, collision avoidance likely imposes a lower bound on streamwise and spanwise proximity, while visual communication may favor positions with better visibility of other flock members or external cues. The model should therefore be interpreted as an aerodynamic baseline within a larger multi-constraint biological decision landscape. Moreover, birds likely do not solve a global energetic optimization problem in any explicit cognitive sense, and it remains unclear whether real birds directly prioritize some recognizable aerodynamic or energetic quantity \cite{bishop2024flying, pennycuick1969mechanics, sachs2012flying, johansson2024aerodynamic}, as is encapsulated in the current cost function formulation.

Taken together, these simplifications may contribute to quantitative discrepancies with field measurements, such as the reported difference in predicted streamwise spacing, and likely lead to some overestimation of wake benefits and/or the role of these wake benefits in determining avian formation structures. Direct assessment of the quantitative impacts of these approximations is challenging, as there does not yet exist a more complex version of this model against which to compare and empirical data on the flapping kinematics of birds in group formations are extremely sparse. Rather, our model's agreement with the experimental observations that do exist may be taken as a signal that these approximations are in some sense reasonable, at least to first order -- but introducing further complexity, both in the physical model itself as well as via additional terms which directly capture biological or social drivers, may well lead to even higher-fidelity results. This presents an exciting line of future research.

However, even with these limitations, this new framework fills a methodological gap between overly simplistic fixed-wing analogies and inaccessibly complex high-fidelity simulations.  We provide a tractable, physics-based model that connects wake structure, force fields, energetic costs, and wing kinematics -- thus establishing a platform for future extensions that can incorporate finer temporal resolution within the stroke, more realistic wake evolution and decay, body and viscous effects, direct prediction of optimal flight speed, or even behavioral flock dynamics such as leader switching. The model presented here may thus be viewed as a minimal, foundational framework: it captures the essential physics of a two-bird aerodynamic interaction, as evidenced by its agreement with empirical observations, and can be readily extended to address more complex aspects of formation flight, such as the evolution of optimal configurations and flight speeds over time, the organization and stability of multi-tiered flocks, and parametric comparisons across avian species.

\matmethods{Here we develop necessary details to model the total force generated by the leader's vortical wake on the follower. We develop an approximate form for this force for an arbitrary leader-follower configuration, as described in Fig. \ref{fig:wake_structure}. The follower is modeled as a lifting line which is oriented in the spanwise $\hat{y}$ direction, has fixed streamwise and vertical coordinates $x^*$ and $z^*$, and is endowed with some circulation distribution. In principle, invoking the Kutta-Joukowsky theorem \cite{cummings2015applied,bai2014generalized} and integrating along the span in a blade-element sense, this wake-induced force with respect to a given leader-follower up/down configuration may be written as
\begin{equation}\label{eqn:full_integral}
    \vec{F}_{\rm{uu,ud,du,dd}} = \int_{y^-}^{y^+} \rho \vec{v}_{\rm{uu,ud,du,dd}}(y;x^*,z^*)\times \vec{\Gamma}(y;x^*,z^*) dy
\end{equation}
where the integral is taken across the span of the lifting line. Here, $\vec{v}_{\rm{uu,ud,du,dd}}(y;x^*,z^*)$ is the wake-induced velocity in the $(x,z)$ plane at a given point $(x^*,y,z^*)$ along the span, and $\vec{\Gamma}(y;x^*,z^*)$ encodes the follower's spanwise circulation distribution. For computational tractability during the optimization procedure, we approximate this integral via a trapezoidal approximation with a single interval: 

\begin{equation}
    \vec{F}\approx \frac{\rho \vec{v}(y^+)\times \vec{\Gamma}(y^+) + \rho \vec{v}(y^-)\times \vec{\Gamma}(y^-)}{2}(y^+-y^-),
\end{equation}
where the $x^*$ and $z^*$ labels, as well as the up/down subscripts, have been dropped for ease of notation. This choice of integral approximation is made to maximize the model's tractability, and to allow for the extraction of the closed-form analytical results developed in the work. Introducing more than two nodes into this trapezoidal approximation would allow for the encoding of arbitrary spanwise circulation distributions into the model, which could further increase the model's accuracy while also increasing its computational requirements. This presents an interesting tradeoff to explore in a future iteration of the model.

We further make the simplifying assumption that $\vec{\Gamma}(y^{\pm})=\Gamma \hat{y}$. That is, we assume that at both of the follower's wingtips, the local circulation is equal to the characteristic circulation $\Gamma$ which defines the bird's vortical wake. This yields
\begin{equation}\label{eqn:integral_approx}
    \vec{F}\approx \frac{\rho\Gamma(y^+-y^-)}{2}\left(\vec{v}(y^+)+\vec{v}(y^-)\right)\times\hat{y}.
\end{equation}
Using Eq. (\ref{eqn:integral_approx}), each force contribution $\vec{F}_{\rm{uu}}$, $\vec{F}_{\rm{ud}}$, $\vec{F}_{\rm{du}}$, and $\vec{F}_{\rm{dd}}$ may be calculated and substituted into Eq. (\ref{eqn:total_induced_force}) to yield the total time-averaged wake-induced force on the follower bird. It just remains to define $\vec{v}_{\rm{uu,ud,du,dd}}(y^{\pm};x^*,z^*)$.

\subsection*{Calculation of induced flow velocity}

Here we define $\vec{v}_{\rm{uu,ud,du,dd}}(y^{\pm};x^*,z^*)$, the flow induced by the leader's vortical wake at the follower's wingtips in a given leader-follower configuration, in order to make use of Eq. (\ref{eqn:integral_approx}). The leader's wake comprises 8 finite-length Rankine vortex filaments, as shown in Fig. \ref{fig:wake_structure}. The contributions from each of these filaments are linearly combined (again, dropping the up/down subscripts):
\begin{equation}
    \vec{v}(y^{\pm};x^*,z^*) = \sum_{i=1}^8 \vec{v}_i(y^{\pm};x^*,z^*).
\end{equation}

A given vortex filament $i$, which has signed circulation $\Gamma$, is characterized by endpoints $\vec{r}_{i,1}$ and $\vec{r}_{i,2}$, with positive direction pointing from $\vec{r}_{i,1}$ to $\vec{r}_{i,2}$. To avoid divergence to infinity, these vortex filaments are endowed with Rankine cores, i.e., cores which undergo solid-body rotation. The radius of the core is $r_c=0.171b$ \cite{Young2003VortexCore}, where $b$ is the semispan length.

Consider a point in space $\vec{r}_0 = (x^*, y^{\pm}, z^*)$ and make the definitions  
\begin{align}
    \vec{a}_i &= \vec{r}_{i,2}-\vec{r}_{i,1} \\
    \vec{b}_i &= \vec{r}_0-\vec{r}_{i,1} \\
    \vec{c}_i &= \vec{r}_0 - \vec{r}_{i,2}.
\end{align}
Then, the orthogonal distance from $\vec{r}_0$ to the vortex filament $i$ is
\begin{equation}
    r_i = \frac{|(\vec{r}_0-\vec{r}_{i,1})\times(\vec{r}_0-\vec{r}_{i,2})|}{|\vec{r}_{i,2}-\vec{r}_{i,1}|} = \frac{|\vec{b}_i\times\vec{c}_i|}{|\vec{a}_i|}.
\end{equation}

Away from the vortex filament's core, i.e., for $r_i>r_c$, the velocity induced at $\vec{r}_0$ is given by \citep{bertin1998aerodynamics}
\begin{equation}\label{v_ind}
    \vec{v}_i(\vec{r}_0) = \frac{\Gamma}{4\pi}\frac{\vec{b}_i\times\vec{c}_i}{|\vec{b}_i\times\vec{c}_i|^2}\left[\vec{a}_i\cdot\left(\frac{\vec{b}_i}{|\vec{b}_i|}-\frac{\vec{c}_i}{|\vec{c}_i|}\right)\right].
\end{equation}

Within some small radius $r_i\leq r_c$ near the $i$th filament's center, the induced velocity may (multiplying (\ref{v_ind}) by $r_i^2/r_c^2$) be written as

\begin{align}
    \vec{v}_i(\vec{r}_0) = \frac{\Gamma}{4\pi}\frac{\vec{b}_i\times\vec{c}_i}{r_c^2}\left[\frac{\vec{a}_i}{|\vec{a}_i|^2}\cdot\left(\frac{\vec{b}_i}{|\vec{b}_i|}-\frac{\vec{c}_i}{|\vec{c}_i|}\right)\right].
\end{align}

It is straightforward to show that this piecewise definition of $\vec{v}_i$ is continuous and finite for all $\vec{r}_0$. It just remains now to define $\vec{r}_0$, $\vec{r}_{i,1}$, and $\vec{r}_{i,2}$ in the context of the available leader and follower configurations.

\subsubsection*{Follower ``up'' and ``down'' configurations}

The follower's reference coordinates are $(x_F,y_F,z_F)$. In its ``up'' configuration (Figs. \ref{fig:wake_structure}(a,c)), the follower's wingtips lift in the $z$ direction by $h_F/2$, so that the wingtip positions are given by
\begin{equation}
    \vec{r}_0 = (x_F, y_F\pm b R_F, z_F + h_F/2).
\end{equation}
In the context of Eqs. (\ref{eqn:full_integral}-\ref{eqn:integral_approx}), this yields $x^*=x_F$, $y^{\pm}=y_F\pm b R_F$, and $z^*=z_F+h_F/2$. The contraction factor $R_F$ is specific to the upstroke.

In the follower's ``down'' configuration (Figs. \ref{fig:wake_structure}(b,d)), its wingtips lower in the $z$ direction by $h_F/2$ so that 
\begin{equation}
    \vec{r}_0 = (x_F, y_F\pm b , z_F - h_F/2).
\end{equation}
Now, in the context of Eqs. (\ref{eqn:full_integral}-\ref{eqn:integral_approx}), this yields $x^*=x_F$, $y^{\pm}=y_F\pm b $, and $z^*=z_F-h_F/2$. There is no contraction factor present in the downstroke configuration, as the follower's wings are fully extended.

\subsubsection*{Leader ``up'' and ``down'' configurations}
The leader's reference coordinates are $(x_L,y_L,z_L)$. In its ``up'' configuration (Figs. \ref{fig:wake_structure}(a,b)), the leader's trailing wake takes the form of an upstroke-generated rectangular array of 4 finite-length vortex filaments immediately behind the leader, as well as a downstroke-generated rectangular array of 4 finite-length vortex filaments located a distance of $\lambda/2$ downstream. The upstroke-generated component of the wake involves tip vortices in the $(x,z)$ plane defined by
\begin{align}
    \vec{r}_{1,1} &= (x_L,y_L+ \pi bR_L/4,z_L+h_L/2) \\
    \vec{r}_{1,2} &= (x_L+\lambda/2,y_L+ \pi bR_L/4,z_L-h_L/2)
\end{align}
and
\begin{align}
    \vec{r}_{2,1} &= (x_L,y_L- \pi bR_L/4,z_L+h_L/2) \\
    \vec{r}_{2,1} &= (x_L+\lambda/2,y_L- \pi bR_L/4,z_L-h_L/2),
\end{align}
where the factor $\pi/4$ is associated with wingtip roll-up \citep{hummel1983aerodynamic, hainsworth1988induced}. The starting vortex associated with the upstroke wake is defined by
\begin{align}
    \vec{r}_{3,1} &= (x_L,y_L- \pi bR_L/4,z_L+h_L/2) \\
    \vec{r}_{3,2} &= (x_L,y_L+\pi bR_L/4,z_L+h_L/2),
\end{align}
and the stopping vortex associated with the upstroke wake is defined by
\begin{align}
    \vec{r}_{4,1} &= (x_L+\lambda/2,y_L- \pi bR_L/4,z_L-h_L/2) \\
    \vec{r}_{4,2} &= (x_L+\lambda/2,y_L+\pi bR_L/4,z_L-h_L/2).
\end{align}

For the downstroke-generated component of the wake, which lies $\lambda/2$ downstream from the leader, the tip vortex filaments are defined by
\begin{align}
    \vec{r}_{5,1} &= (x_L+\lambda/2,y_L+ \pi b /4,z_L-h_L/2) \\
    \vec{r}_{5,2} &= (x_L+\lambda,y_L+ \pi b /4,z_L+h_L/2)
\end{align}
and
\begin{align}
    \vec{r}_{6,1} &= (x_L+\lambda/2,y_L- \pi b /4,z_L-h_L/2) \\
    \vec{r}_{6,2} &= (x_L+\lambda,y_L- \pi b /4,z_L+h_L/2).
\end{align}
The starting vortex associated with the downstroke wake is defined by
\begin{align}
    \vec{r}_{7,1} &= (x_L+\lambda/2,y_L- \pi b /4,z_L-h_L/2) \\
    \vec{r}_{7,2} &= (x_L+\lambda/2,y_L+\pi b /4,z_L-h_L/2),
\end{align}
and the stopping vortex associated with the downstroke wake is defined by
\begin{align}
    \vec{r}_{8,1} &= (x_L+\lambda,y_L- \pi b /4,z_L+h_L/2) \\
    \vec{r}_{8,2} &= (x_L+\lambda,y_L+\pi b /4,z_L+h_L/2).
\end{align}
To define the wake structure for the leader's ``down'' configuration, a similar procedure as above is carried out. The upstroke portion is shifted in the positive streamwise direction by $\lambda/2$, and the downstroke portion is shifted in the negative streamwise direction by $\lambda/2$, so that the two structures exchange places. 

\subsection*{Details of implementation of optimization procedure}
We define the optimization problem, which is solved numerically in this work, using the built-in \textit{optimproblem} framework in MATLAB. The six optimization variables are defined symbolically as \textit{optimvar}s. The \textit{fmincon} solver option is then passed into \textit{optimoptions}, with the algorithm chosen to be Sequential Quadratic Programming (SQP). This particular choice of algorithm is standard for problems in which the objective function and constraints are nonlinear while still continuously differentiable \cite{boggs1995sequential}. All other parameters are left as their MATLAB defaults. 

}

\showmatmethods{} % Display the Materials and Methods section

\acknow{This work was supported by the Hope Street Postdoctoral Fellowship at Brown University,  ONR Grant N00014-21-1-2816 and NSF Award IOS-1930924.}

\showacknow{} % Display the acknowledgments section

\bibliography{pnas-sample}

\end{document}